# Observation of the Meissner state in superconducting arrays of 4-Ångstrom carbon nanotubes

*Chao Ieong, Zhe Wang, Wu Shi, Yuxing Wang, Ning Wang, Zikang Tang,
Ping Sheng, and Rolf Lortz[*]*

Department of Physics and William Mong Institute of Nano Science and Technology   Hong Kong University of Science and Technology
Clear Water Bay, Kowloon, Hong Kong, China

## Abstract

We report clear observations of the magnetic Meissner Effect in arrays of superconducting 4 Å carbon nanotubes grown in the linear channels of $AlPO_4$-5 (AFI) zeolite single crystals. Both bulk magnetization and magnetic torque experiments show a pronounced signature of the lower critical $H_{c1}$ transition, a difference in zero-field cooled and field cooled branches during temperature sweeps below 6K, and signatures of 1D superconducting fluctuations below ~15-18 K. These experiments extend the magnetic phase diagram we obtained previously by resistive experiments [Z. Wang et al., Phys. Rev. B **81**, 174530 (2010)] towards low magnetic fields and within the range of zero resistance.

**PACS** numbers: 73.63.Fg, 74.70.Wz, 74.25.Ha, 74.81.Fa

## I. INTRODUCTION

The observation of superconductivity among the rich variety of known structures of elementary carbon was considered for a long time as limited to $C^{60}$ buckyballs with interstitial ions [1], or boron-doped diamond [2] and in various intercalated graphite compounds [3]. Due to the high symmetry of the graphene sheet, the electron-phonon coupling - the most important parameter for phonon-mediated superconductivity - is particularly weak. However, it has been predicted theoretically that the electron-phonon coupling can be enhanced significantly by the large curvature of small carbon nanotubes (CNTs) [4]. A superconducting ground state within a quasi-1D CNT would be in competition with thermal fluctuations [5,6] and possibly the insulating ground state related to the Peierls distortion [7]. Nevertheless, recently we reported solid evidences for superconductivity with electrical resistive transition in arrays of 4 Å CNTs (CNT@AFI), which we grow in the linear pores of AFI zeolite single crystals (composition: $Al_{12}P_{12}O_{48}$) [8-11].

Early observations of superconductivity in this composite material on the nanoscale were limited to a broad continuous decrease of the bulk magnetization and the observation of a supercurrent gap feature in the current-voltage (*IV*) characteristics below ~15 K. Recently, we have been able to fabricate samples with higher carbon filling factor of the AFI pores. With these samples we were able to probe true zero resistivity in the superconducting state, details of the supercurrent gap [9] and the thermodynamic signature of the superconducting transition in the specific heat [10]. These new samples provide now access to much more detailed investigations of the superconducting properties of this rather intriguing material comprising arrays of ultra-thin, linear and parallel metallic CNTs embedded in an insulating host material.

---
[*] Corresponding author: lortz@ust.hk

Such arrays of weakly transversally coupled superconducting CNTs belong to the family of quasi-one-dimensional superconductors, such as $Tl_2Mo_6Se_6$ (which is related to the Chevrel phase superconductors [12]) or some organic superconductors as e.g. pressurized $(TMTSF)_2PF_6$ [13]. The highly anisotropic nature of such superconductors may imply the formation of intrinsic arrays of Josephson junctions (JJ) [14,15] within their crystalline structure. The global superconducting transition of arrays of JJs has been demonstrated to fall in the same universality class as the Berezinski-Kosterlitz-Thouless (BKT) transition [16]. In two previous papers we could show for both, CNT@AFI [9] and $Tl_2Mo_6Se_6$ [17], that longitudinal phase slips (which in a truly 1D superconductor always lower the temperature of global phase coherence down to zero temperature [5]) are strongly suppressed below a phase-ordering transition in the lateral plane to the 1D direction, once a significant coupling of the 1D chains / CNTs via the Josephson Effect exists. This transition exhibits the characteristics of a BKT-like transition, indicating that we can model such quasi 1D superconductors indeed as intrinsic arrays of Josephson junctions.

In CNT@AFI, superconducting fluctuations with 1D character were observed below ~15 K in the form of a gradual decrease of resistivity, a gap-like feature in the *IV*-characteristics and the onset of the broad specific-heat transition [9,10]. In samples with higher filling factor of the AFI pores, the resistivity (probed on a length scale of only ~100 nm) shows reproducibly an additional jump to zero resistivity below 7 K [9]. Complex *IV* characteristics were observed with several plateaus in the differential resistivity which have been interpreted as fingerprints of a BKT-like transition for which theory predicts nonlinear *IV* characteristics with $I \sim V^\alpha$. A complex critical-field transition is furthermore found in magnetic fields: Zero resistivity occurs at low temperatures up to applied fields of ~3 T. In higher fields, the resistivity increases due to the gradual loss of 3D phase coherence associated with the increasing strength of field-induced fluctuations in the phase of the order parameter. The resistivity shows a pronounced anisotropy for field orientations parallel and perpendicular to the CNTs which vanishes in a field of ~18 T where the resistivity tends to saturate. However, a small finite slope indicates the persistence of superconducting correlations up to 28 T [9].

Electrical transport experiments provided us with a wealth of information on the highly complex nature of the superconducting transition in CNT@AFI. Nevertheless, they left an open question, whether a true Meissner state and Abrikosov state can be distinguished within the range of zero resistivity. Magnetization experiments are the most direct method for the investigation of the Meissner state of superconductors. However, in the present case, such experiments — especially in small magnetic fields — represent a particular challenge owing to the weakness of the superconducting signal compared to the paramagnetic and diamagnetic background contributions from the insulating zeolite host, and possibly also from some unreacted carbon. In order to overcome this difficulty, we used two different experimental approaches: In this paper, measurements of both high-resolution bulk magnetization and magnetic torque are presented. The latter quantity probes the anisotropic component of the magnetization, related to the 1D nature of the CNTs, and is expected to provide a larger signal to background ratio as compared to the bulk magnetization. For both type of measurements, we report clear signatures of a Meissner transition in low magnetic fields as well as traces of superconducting fluctuations up to high magnetic fields.

## II. SAMPLE PREPARATION AND EXPERIMENTAL DETAILS

AFI zeolite crystals (~100 μm in diameter and ~500 μm long) containing the tripropylamine precursor in their straight pores (aligned along the crystalline $c$- axis with a center-to-center separation of 1.37 nm) were first heated at 580 °C in 0.7 atm of nitrogen and 0.3 atm of oxygen for 8 h to remove the precursor tripropylamine in the linear pores, and then in 0.7 atm nitrogen and 0.3 atm of ethylene at the same temperature for 8 h, as described in detail elsewhere [9]. The diameter of the pores, after discounting the size of oxygen atoms lining the walls, is 0.7 nm [18]. The typical size of a single crystal is 0.3 mm x 0.05 x 0.05 mm with mass of a few tens of micrograms.

The bulk magnetization was measured with a commercial Quantum Design™ VSM SQUID magnetometer. We selected ~ 50 single crystals from the same batch as in Ref. [10] and aligned them parallel to each other on a Quartz sample holder of the VSM SQUID and fixed them with vacuum grease (which is slightly diamagnetic). The measurements were performed for field alignment both perpendicular and parallel to the $c$-axis of the CNTs. This allowed us to derive the anisotropic component of the magnetization $M_{aniso.}=M_{parallel}–M_{perpendicular}$. This anisotropic component of the bulk magnetization is rather insensitive to the mostly isotropic paramagnetic and diamagnetic contributions of the zeolite and the grease and therefore useful to look for weak signals related to superconducting fluctuations of the CNTs above $T_c$ or in high magnetic fields.

The magnetic torque was measured with a homemade capacitive cantilever technique. Two parallel cantilevers made copper-beryllium foil were mounted parallel to each other so that their active areas form the two plates of a capacitor ($C$~10 pF). A 50 μm thin sapphire disk was used to electrically isolate the legs of the cantilevers from each other. About 50 CNT@AFI crystals were aligned parallel to each other on one of the round plates of the cantilevers of the torque sensor and fixed with vacuum grease. A manual analogue General Radio Capacitance Bridge in combination with a modern digital lock-in amplifier was used to determine relative changes in the separation of the cantilever plates due to the magnetic torque of the sample with high resolution. The construction of the torque sensor allows us to perform highly reversible measurements during both temperature and field sweeps. The analysis of the data taken during temperature sweeps requires a correction of the contribution from the thermal expansion of the sapphire spacer and the capacitor plates, which was carefully determined in a separate measurement in zero field. The samples were aligned so that the $c$-axis of the CNTs was oriented 45 degrees with respect to the magnetic field to ensure a large signal. The magnetic torque is defined as $\boldsymbol{\tau} = \boldsymbol{M} \times \boldsymbol{H}$, where $\boldsymbol{M}$ is the anisotropic component of the magnetization and $\boldsymbol{H}$ the applied magnetic field.

## III.  EXPERIMENTAL RESULTS

### 3A.  Magnetic torque measurements

In Fig. 1 we plot the magnetic torque measured in a magnetic field of 100 G during a temperature sweep after zero-field cooling (ZFC) and field cooling (FC). At 6 K a pronounced downturn occurs with a clear difference between the ZFC and FC curves. The ZFC data shows a larger torque magnitude of negative signature. The two branches meet at 6 K at the onset of the downturn. The negative signature at low temperature indicates that the sample attempts to align the c-axis of the hexagonal AFI crystals (which is also the direction of the *c*-axis of the CNTs) parallel to the applied field.

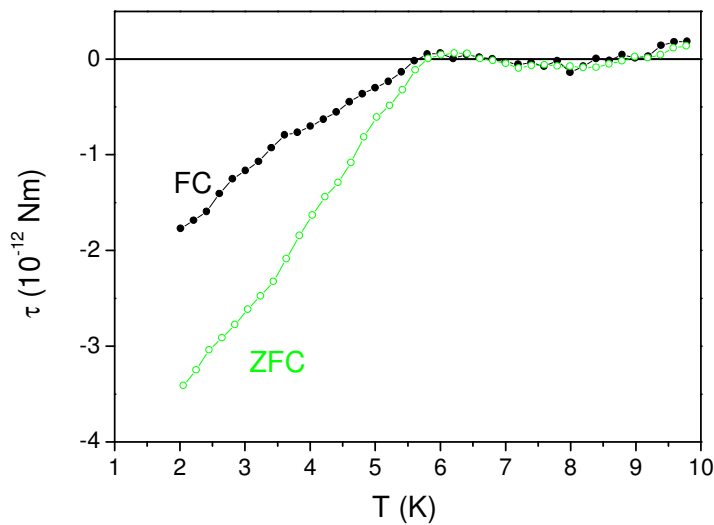

**Figure 1.** Variation of the magnetic torque signal of the 4 Ångstrom carbon nanotubes embedded in AlPO$_4$-5 zeolite single crystals during heating in a field of 100 G after zero-field cooling (green, ZFC) and field cooling (black, FC).

The inset of Fig. 2 shows the raw data obtained from our torque sensor at fixed temperatures $T = 1.7$ K and 20 K measured during field sweeps up to 10 T. The sample had been cooled in zero field before the measurements. Data have been recorded during slow field sweeps from 0 to +10 T to -2 T and back to +10 T. We have chosen a logarithmic field scale to illustrate the behavior in small fields more clearly. Under magnetic fields larger than ~0.1 T only a tiny difference is found between the two data sets. The strong field-dependent background mostly originates from eddy currents which lead to attractive forces between the capacitor plates of the cantilever. As long as the field-sweep rates are kept small, this background is very reproducible and temperature independent. In order to extract the superconducting signal, we derive the difference $\Delta\tau$ between the two curves (see main Figure). In fields of less than 3 T the toque signal is nearly constant and a sharp downturn appears below 0.05 T. The ZFC initial curve clearly differs from the following data owing to the hysteretic behavior detailed below. Above ~3 T the torque signal slowly approaches zero.

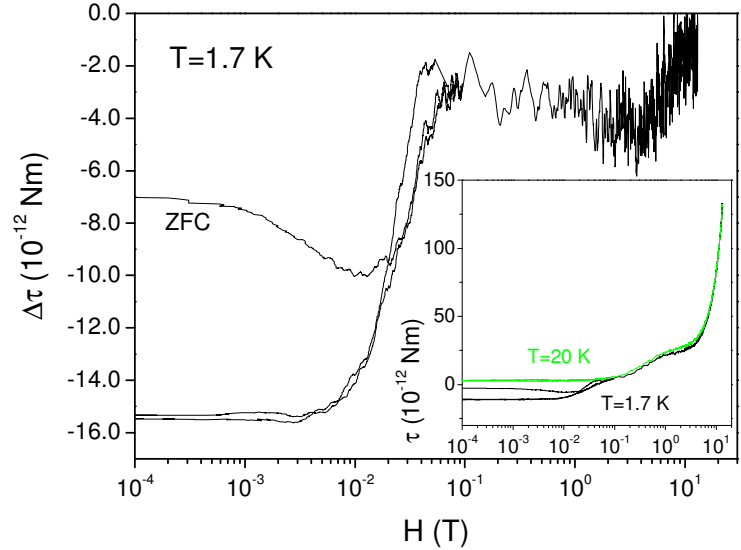

**Figure 2.** Inset: Magnetic field dependence of the total magnetic torque signal of the 4 Ångstrom carbon nanotubes embedded in AlPO$_4$-5 zeolite single crystals at 1.7 K (green) and 20 K (black). The main figure shows the difference $\Delta\tau$ in the two curves (with the green curve subtracted off) which represents the (hysteretic) superconducting contribution to the torque.

Figure 3 shows the temperature dependence of the magnetic torque under low magnetic fields in the original data on a linear field scale. At 16 K the torque signal shows only little field dependence. A dip in the torque signal appears at temperatures below 6 K in fields smaller than 0.04 T, which becomes very pronounced at the lowest temperatures measured. Hysteretic behavior, with a clearly different initial zero-field cooled curve, is visible below 4.5 K. The initial curve starts from zero field as an approximately straight line and then passes through a minimum and increases in higher fields. This is the typical behavior for the reversible magnetization of a type II superconductor close to the lower critical field $H_{c1}$. $H_{c1}$ is generally defined as the field of first flux penetration into the sample and therefore indicated by the point where the magnetization starts to deviate from linear behavior (here at ~60 Oe). When the field is swept to high value and then back to zero, irreversibility sets in and the torque remains at a finite value. We attribute this residual signal to some trapped flux in the sample which interacts with either a small component of a non-axial residual magnetic field of our superconducting solenoid, or the earth's magnetic field.

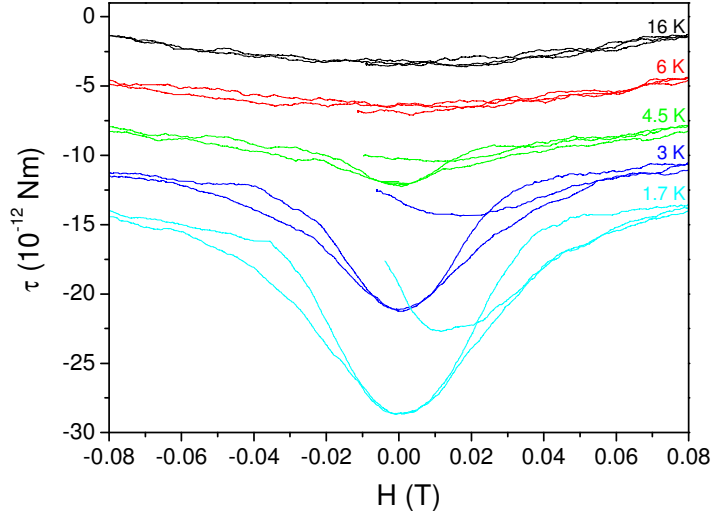

**Figure 3.** Magnetic field dependence of the total magnetic torque signal of the 4 Ångstrom carbon nanotubes embedded in AlPO$_4$-5 zeolite single crystals at various temperatures showing the temperature evolution of the superconducting Meissner state.

## 3B. Discussion

A significant torque signal of negative signature appears sharply at 6 K where the ZFC and FC data start to deviate from each other. A resistive transition below 6 K was observed in the same temperature range [9], below which global 3D phase coherent superconductivity with macroscopic screening currents is established. As we have a system of 1D parallel CNTs, zero resistivity and a well-defined Meissner state can only occur if a significant coupling of the CNTs via the Josephson Effect exists [5]. In a phase-coherent bulk 3D superconductor these screening currents can flow freely and the long hexagonal *c*-axis of the AFI crystals will have a strong tendency to align parallel to the applied field in order to reduce the demagnetization factor (Fig. 4a). This is in accordance with our negative torque signal. The larger magnitude of the ZFC data (Fig. 1) further proves the existence of macroscopic superconducting screening currents which can only flow in a bulk superconducting state and not in individual uncoupled 1D CNTs with superconducting correlations.

### a) 3D Bulk Superconducting State

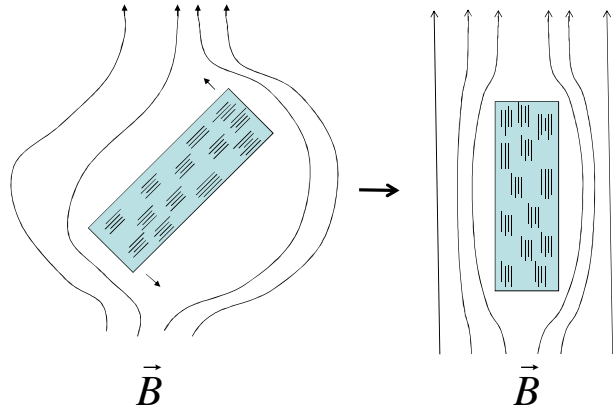

**Figure 4a)** Cartoon illustrating the origin of the magnetic torque in a phase-coherent 3D bulk superconducting Meissner state formed by superconducting CNTs strongly coupled via the Josephson Effect. The rectangular area represents the AFI crystal and the parallel lines the CNTs in the AFI channels which are grouped into regions of high density. The black arrows illustrate the forces acting on the crystal due to the external applied ***B*** field. The figure on the right represents the state of lower energy as the ***B***-field lines are less bended (smaller demagnetization factor).

### b) 1D Superconducting Fluctuations

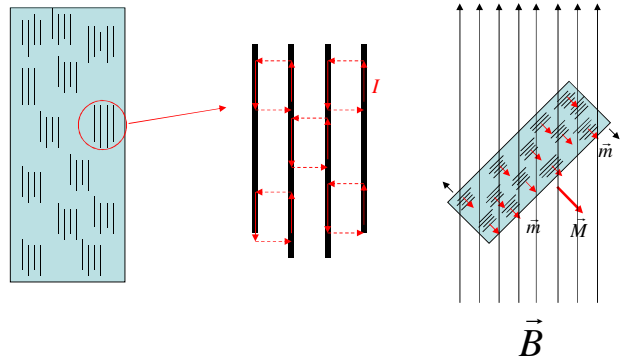

**Figure 4b)** Cartoon illustrating the origin of magnetic torque in a sample with little electronic coupling of the CNTs. Screening currents (red arrows) are then confined into the CNTs nanotubes with tunneling currents between neighbors forming Josephson vortices. This would results in microscopic magnetic moments $\vec{m}$ (small red arrows) perpendicular to the CNTs which sum up to a macroscopic magnetic moment $\vec{M}$ (large red arrow) perpendicular to the *c*-axis of the AFI crystals. The black arrows indicate the forces on the crystal due to the interaction of the magnetic field with the induced diamagnetic moment which attempt to rotate the crystal into the opposite direction as in Fig. 4a. Such samples would not form any true Meissner state and would show a magnetic anisotropy with a sign that is opposite to that in Fig. 4a.

Note that, dependent on its microscopic origin within the CNT@AFI crystals, the magnetic torque signal could have either a positive or negative signature. In our present CNT@AFI samples only 10 ± 2 % of the total pore length is occupied by CNTs [10]. The anisotropy depends therefore on the exact configuration of the CNT arrays inside the zeolite crystals. Most likely, short segments of CNTs are being grouped into regions of high CNT density with emptier regions in between (compare Fig. 4a and b). Dependent on the average density of the CNTs, a change from 1D to 3D superconductivity should occur at a critical filling factor of the pores when the coupling of the superconducting CNTs via the Josephson Effect becomes effective. In a sample with dominating 3D character and a well developed Meissner state, macroscopic screening current will try to align the *c*-axis of the AFI crystals (which is also the c-axis of the CNTs) parallel to the field direction in order to minimize the demagnetization factor (Fig. 4a). However, in a sample with dominating 1D character, screening currents would be confined into the 1D CNTs. Such currents could only flow in the form of Josephson vortices [19] with tunneling currents between neighboring tubes within phase-coherent regions of the sample (Fig. 4b). The latter type of samples should not show any well-developed Meissner state. The induced diamagnetic moment would then point to a direction perpendicular to the CNTs, trying to align the CNTs perpendicular to the applied field. Earlier bulk magnetization experiments on samples with lower carbon content in the pores showed a larger diamagnetic signal for a perpendicular field orientation and a one order of magnitude smaller signal for the parallel field orientation [8]. This would have caused a positive magnetic torque signal in the superconducting state. The present torque result therefore suggests that our new crystals with a higher filling factor of the AFI pores changed their behavior: Early samples had low filling factors and therefore showed only superconducting correlations of 1D character without macroscopic screening currents. Indeed, in Ref. [8] there was no visible difference in the ZFC and FC branches. Furthermore, these samples only showed a continuous drop of the magnetization signal and a continuous opening of a supercurrent gap in the *IV* characteristics, instead of a sharp superconducting transition anomaly. The new samples with improved filling factor however clearly show the presence of macroscopic screening currents below a sharply defined transition temperature. This indicates that bulk 3D superconductivity and a Meissner state is formed below ~6 K in magnetic fields below ~60 - 100 G. An open question is why $T_c$ is then lower, as one would expect that the enhanced electronic coupling in the new crystals should stabilize superconductivity and eventually even raise $T_c$. Resistivity data in Ref. [9] revealed, however, that the transition at 6 K represents only a part of the superconducting transition where the 3D phase coherent superconducting state is triggered by a BKT-like transition in the lateral plane of the CNTs. Superconducting fluctuations were observed up to 15 K, in accordance with the earlier bulk magnetic measurements. The experimental limitations, which are basically determined by the drift of the capacitance measurement, do not allow us to judge whether this small fluctuation component enters the torque signal at higher temperatures. As we will show in the following section, these fluctuations do indeed show up in bulk magnetization measurements on the samples under investigation.

### 3C. Bulk magnetization measurements

Figure 5a shows a bulk magnetization measurement at $T = 1.8$ K during a field sweep. The S-shaped curve is formed by the simultaneous presence of a paramagnetic and a diamagnetic background, which we ascribe mainly to contributions from the zeolite host material. In small fields, a tiny sharp feature is observed as indicated by the circle. Figure 5b shows an enlargement of this area. Hysteretic behavior is observed, consistent with all the characteristics of a superconductor in the vicinity of $H_{c1}$. The initial curve illustrates the characteristic linear behavior of the Meissner state, whereas in what follows the curve indicates some magnetic flux being trapped in the volume of the sample. Note that the curves are point symmetric to the origin of the graph but in case of the magnetic torque $\tau = M \times H$ the *y*-axis represents an axis of mirror symmetry. $H_{c1}$ perfectly agrees with our previous estimate from the magnetic torque measurements.

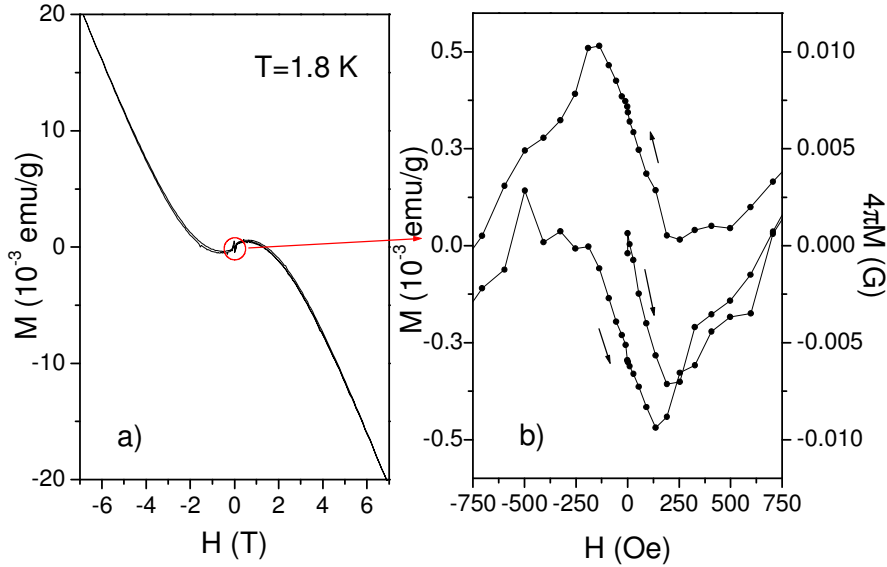

**Figure 5. a)** Magnetization loop at $T = 1.8$ K. The circle encloses the anomaly at $H_{c1}$. **b)** Details in the magnetization in small fields around the superconducting $H_{c1}$ transition. The field has been applied perpendicular to the *c*-axis of the CNTs.

Figure 6 show the temperature dependent magnetization under both ZFC and FC conditions. A clear superconducting signal appears below 9 K in a field of 20 G with a strong difference between the ZFC and FC branches. A magnetic field lowers the temperature of this magnetization anomaly. Zooming in on the onset of the transition (inset) illustrates that superconducting fluctuations start to form in the temperature range of 15 to 18 K. The data perfectly agrees with our previous torque, electric transport [9,10] and specific-heat experiments [10]. Note that the low-field signature of superconductivity shows only a minor anisotropy, indicating the bulk 3D nature of superconductivity in this field range.

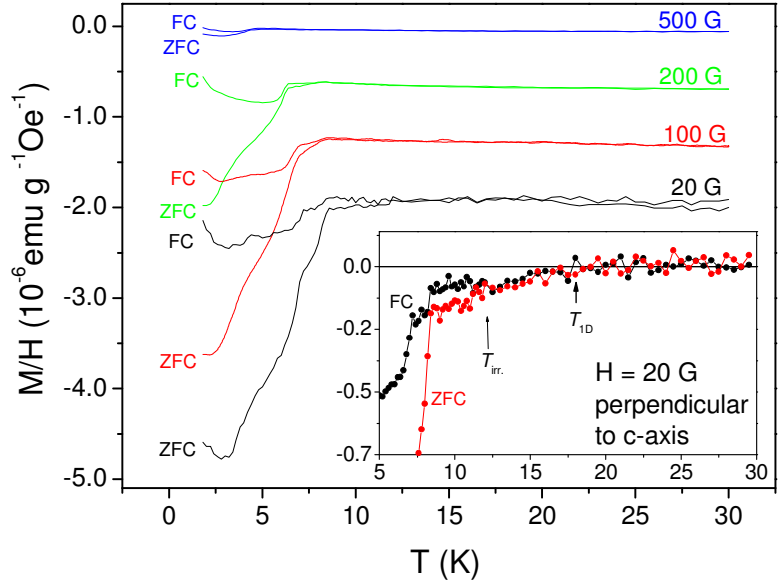

**Figure 6.** Superconducting transition in the bulk magnetization in various small magnetic fields applied perpendicular to the *c*-axis of the CNTs for the ZFC and FC conditions. The curves have been shifted relative to each other for clarity. Inset: Details of the superconducting transition in the bulk magnetization in a field of 20 G applied perpendicular to the *c*-axis of the CNTs (a linear background fitted in the temperature range of 20 - 30 K has been subtracted). The arrows indicate the onset of 1D superconducting fluctuations ($T_{1D}$) and of irreversibility ($T_{irr.}$).

Magnetization data in higher magnetic fields is hampered by the strong temperature-dependent Curie-Weiss term which originates from the paramagnetic background. This background is however largely isotropic. In Figure 7 we plot therefore the anisotropic component $M_{aniso.} = M_{parallel} - M_{perpendicular}$ from magnetization data obtained for magnetic fields aligned parallel and perpendicular to the *c*-axis of the CNTs. The comparably sharp large transition anomalies into the Meissner state, as observed in Figure 6, hardly enter the data in Figure 7. The remaining signal is represented by a broad continuous reduction of the anisotropic magnetic moment. This downturn is also visible in Figure 6, as demonstrated in the inset, however nearly one order of magnitude smaller. The curves reveal an anisotropic contribution to the magnetization which is strongly field-dependent up to 1 T. It persists in higher fields up to 7 T with its onset remaining at temperatures somewhat above 15 K and there hardly depends on the field strength. This illustrates the presence of robust 1D superconducting fluctuations up to much higher fields, in accordance with earlier resistivity data [9].

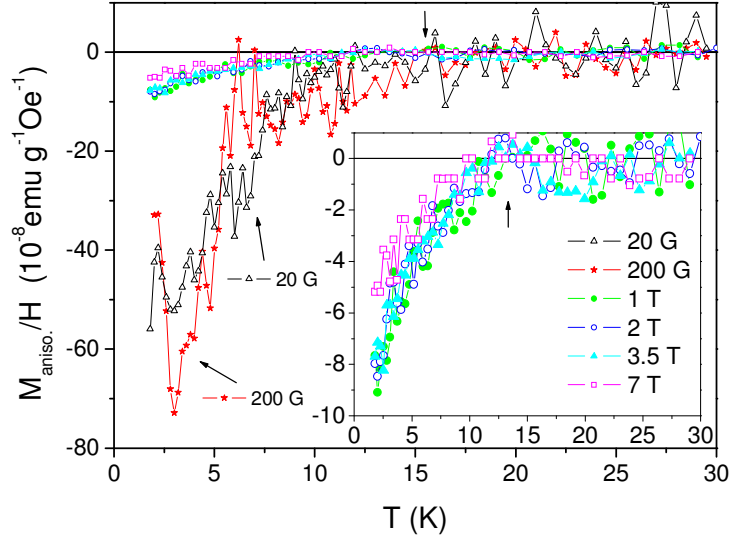

**Figure 7.** Anisotropic component $M_{aniso.}$ of the magnetization in various fields from 20 G to 7 T (see text for details). The arrows indicate the onset temperature of superconducting fluctuations. No background has been subtracted. The inset shows the high-field data (1, 2, 3.5 and 7 T) in more detail.

**3D. Discussion**

Although the initial slope of the data in Figure 5 suggests that we have a true Meissner state established, the diamagnetic screening is far from being perfect. The slope of the initial branch in Figure 5 plotted as the volume magnetization $4\pi M$ in units of G is only $-5 \times 10^{-5}$, whereas $-1$ would be found for a bulk superconductor in the Meissner state. This corroborates with the smallness of the superconducting signal which is about 5 orders of magnitude smaller than that of a typical bulk superconductor of comparable volume. If we take the difference between the ZFC superconducting state and the normal state from the 20 G data in Figure 6, only a slightly larger value of $-1 \times 10^{-4}$ is obtained. The difference can be explained by the slower field sweep rate used during the measurement of the data in Figure 5. The tiny diamagnetic screening value is certainly related to our low filling factor of the pores: We have a composite material on the nanometerscale, formed by CNTs rich regions which are much smaller than the London penetration depth. The magnetic field can therefore penetrate into the volume through regions with low CNT content without disturbing superconductivity within the CNT rich regions too much.

The negative signature of $M_{aniso.}$ demonstrates that superconductivity is more robust in parallel applied fields, or in other words, the diamagnetic moment is of larger magnitude in the parallel field orientation (Fig. 8) with a higher upper critical field transition. This is the typical behavior of bulk anisotropic superconductors and is related to the anisotropy of the Fermi surface, as observed, e.g., in the quasi-1D superconductor $Tl_2Mo_6Se_6$ [12]. If the CNT array shape is like that shown in Fig. 4a, then the state in which the field is parallel to the *c*-axis should be a lower energy state, while the perpendicular field state should be the higher energy state. This is in perfect agreement with the torque data. Hence the magnetization vs. temperature behavior should look like that shown in Fig. 8. That means $M_{parallel}$ should be negative with a larger absolute value.

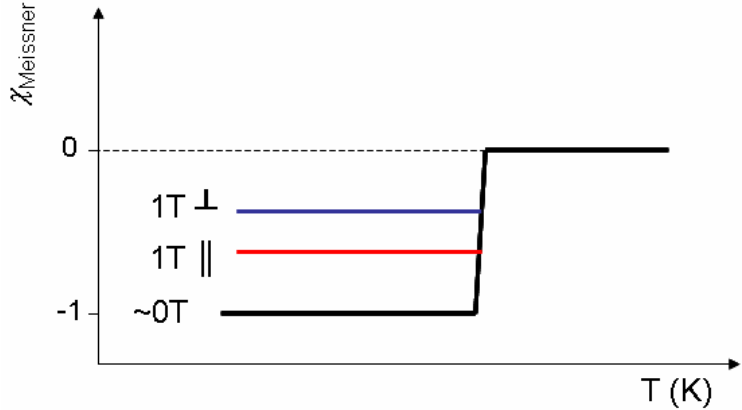

**Figure 8.** Schematic plot of the Meissner magnetization (*M*/*H*) to illustrate the effect of the direction of the applied field in an anisotropic superconductor. The orbital effect on superconductivity is weaker for a field orientation parallel to the *c*-axis of the CNTs due to the more open nature of the Fermi surface. A field applied parallel to the *c*-axis therefore has less influence on the superconducting state and a larger diamagnetic moment results.

Earlier, less sensitive magnetization data [8], showed a larger contribution for perpendicular fields and therefore the opposite behavior. However, it has to be cautioned that the samples used in the present experiment differ from the earlier ones in terms of a higher filling factor of the AFI channels with enhanced electronic coupling of the CNTs. Hence, we expect a difference in the shape of the regions of high CNT density inside the AFI crystals (Fig. 4). This shows the different character of superconductivity related to the higher filling factor of the pores of our AFI samples with strongly enhanced electronic coupling between the CNTs. Furthermore, the earlier data showed larger field effect on superconducting fluctuations in higher magnetic fields, pointing to a significantly lower upper critical field in samples with less coupled CNTs.

The missing field dependence in the data shown in Figure 7 in higher applied fields appears, on a first view, quite unusual and seems to be in contradiction to our earlier specific-heat experiments which showed strong field dependence up to ~3 T. However, resolving the tiny superconducting specific-heat anomaly at $T_c$ required the subtraction of reference data taken in a magnetic field. Due to technical limitations in the accuracy of the field calibration of the calorimeter, data in only 5 T was used as reference in Ref. [10]. From Ref. [9] we know, however, that a field of 5 T only destroys the macroscopic phase coherence whereas 1D superconducting correlations persist up to 28 T. In phase-fluctuation dominated superconductors, the specific heat consists of two contributions: A broad anomaly related to the Cooper pair formation and an anomaly at the actual $T_c$ which represents the phase-ordering transition below which zero resistivity is established. In strongly underdoped cuprate superconductors the onset of Cooper pair formation has been observed at temperatures up to 2 times $T_c$ and the actual phase ordering transition at $T_c$ represented only a small fraction of the total specific heat [20]. This indicates that the observed field-dependent specific-heat contribution represented only a small fraction of the total superconducting contribution and much higher fields would have been required to fully suppress the superconducting 1 D correlations. With $M_{aniso.}$, we are sensitive for the pairing contribution as no subtraction of reference data needed to be done to obtain the data in Fig. 7.

In order to understand the high-magnetic field behavior, let us recall the effect of a magnetic field on a low dimensional superconductor. For 1D superconducting correlations, the orbital limit for superconductivity at which the screening currents around vortex cores reach a Cooper pair breaking value [21] is replaced by the Pauli limit for superconductivity [22,23]. Macroscopic screening currents can not form in a 1D object with open Fermi surface and the only effect of a magnetic field is then the Zeeman splitting of the Fermi energy which reaches a Cooper pair breaking value at the Pauli limit. Typically, such superconductors show only little field dependence for fields smaller than the Pauli limiting field [24] and an abrupt first-order $H_{c2}$ transition once the Pauli limit is reached.

## IV.  PHYSICAL INTERPRETATION AND SUMMARY

In Fig. 9 we summarize our data in the form of a low-magnetic field phase diagram of superconductivity in CNT@AFI. The dashed red line, which starts almost vertically at 15 K, illustrates qualitatively the temperature below which the first signature of diamagnetism is found in the bulk magnetization. At the same temperature we observed previously the onset of the resistive transition and the onset of the specific-heat transition, as well as the opening of a supercurrent gap in the $IV$ characteristics [8-10].

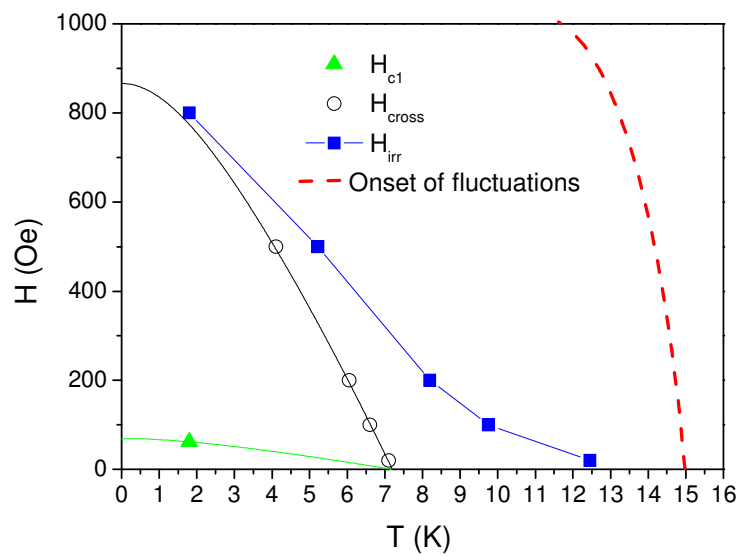

**Figure 9.** Low-magnetic field phase diagram with the lower critical field transition at $H_{c1}$, the critical field $H_{cross.}$ which represents the crossover field from 3D to 1D superconductivity, the onset of flux pinning irreversibility at $H_{irr}$ and the onset of 1D superconducting fluctuations (see text for details).

The triangle represents the lower critical field ($H_{c1}$) transition at $T$=1.8 K, as extracted from the ZFC data shown in Figure 5. The lowest magnetic field where a deviation from linear dependence is observed is chosen. This field should correspond to the field where the first magnetic flux quanta enter the superconducting volume. A clear determination is only possible at the lowest temperature as the $H_{c1}$ line is getting quickly broadened into a crossover upon increasing temperature. A fit with the standard Werthamer-Helfand-Hohenberg (WHH) theory (green line) [25] leads to an estimate of $H_{c1}(0) = 70$ Oe. The presence of zero resistance up to ~ 3 T and the observation of superconducting fluctuations up to 28 T [9] suggest that we have an extremely high-$\kappa$ superconductor.

The rapid growth of the diamagnetic signal below ~9 K is related to the establishment of global 3D phase coherence through the Josephson coupling between the nearest neighbor CNTs in the lateral plane to the CNTs. We considered the temperature of the steepest slope in the zero-field cooled branch as criteria to define this characteristic field ($H_{cross.}$) in the phase diagram. This strong diamagnetic signal vanishes rather rapidly in higher magnetic fields. A fit with the WHH model leads to an estimate of $H_{cross.}(0) = 900$ Oe.

The blue squares represent the temperatures where the zero-field and field cooled branches start to deviate significantly. This defines the irreversibility line below which transversal superconducting phase fluctuations become so much stabilized that some pinning of magnetic flux is observed.

We attribute the region between $H_{c1}$ and $H_{cross.}$ to a region where the system behaves as a bulk type-II superconductor in the Abrikosov state. The region between $H_{cross.}$ and 3 T (where we still observe zero resistance over a length scale of 100 nm [9]), represents an intermediate region where the crossover from 3D bulk superconductivity to 1D superconducting correlations occurs. This crossover originates from a gradual loss of phase coherence in the lateral plane. The observation of zero resistance may therefore depend on the separation of the voltage leads in the experiment (100 nm in Ref. [9]). The applied magnetic field introduces a magnetic length scale into the system, which is closely related to the mean distance between field-introduced vortices penetrating the system above $H_{c1}$ [26,27]. This length scale limits the range of transversal phase coherence and perturbs the longitudinal phase coherence due to the reduced Josephson interaction of neighboring CNTs. In fields beyond 3 T the longitudinal coherence length becomes shorter than the length over which we probed by resistivity and phase slips of the order parameter cause non-zero longitudinal resistance.

Resistivity and specific heat both showed an onset of fluctuations with presumably strongly 1 D character at ~15 K. Although our magnetic torque sensor was not precise enough to resolve this contribution, we demonstrated its presence clearly with help of our anisotropic bulk magnetization data which provided higher accuracy (Figure 7).

## V. CONCLUDING REMARKS

Our magnetic measurements confirm the rather complex nature of superconductivity in this quasi-1D superconductor, which we observed earlier in electric transport experiments. The temperature- and magnetic-field-induced crossover from 3D bulk superconductivity to 1D superconducting fluctuations has been demonstrated to be triggered by a phase-ordering transition among neighboring superconducting CNTs within the lateral plane to the individually fluctuating CNTs, closely related to the Berezinskii-Kosterlitz-Thoules transition [9,17]. A very similar behavior has been observed in the extreme quasi-1D superconductor $Tl_2Mo_6Se_6$ [17] which shows that we can consider such quasi-1D superconductors as a new universality class of superconductors with superconducting properties that are strongly governed by the formation of an intrinsic Josephson junction array within their crystalline structure. Presently, our composite samples are far from being perfect. The investigated samples had filling factors of the AFI pores of ~10 % of the pores. We are currently making a large effort on improving the carbon filling factor (up to ~25%) in order to prepare samples, which can be investigated by more sophisticated methods such as, e.g., scanning tunneling spectroscopy.

## ACKNOWLEDGEMENTS

This work was supported by the Research Grants Council of Hong Kong Grants DAG08/09.SC01, HKUST9/CRF/08, and HKUST 603108.